\documentclass[pss,fleqn]{w-art}
\usepackage{times}
\usepackage{w-thm}
\usepackage{subfigure}  
\usepackage[]{graphicx}
\begin{document}
\DOIsuffix{theDOIsuffix}
\Volume{XX}
\Issue{1}
\Month{01}
\Year{2003}
\pagespan{1}{}
\Receiveddate{15 November 2003}
\Reviseddate{30 November 2003}
\Accepteddate{2 December 2003}
\Dateposted{3 December 2003}



\title[IEC in DMS ML]{Interlayer exchange coupling in (Ga,Mn)As based multilayers}


\author[Giddings]{A.D. Giddings\footnote{Corresponding
     author: e-mail: {\sf ppxadg@nottingham.ac.uk}}\inst{1}}
\address[\inst{1}]{School of Physics and Astronomy, University of Nottingham, Nottingham NG7 2RD, UK}
\author[Jungwirth]{T. Jungwirth\inst{2,1}}
\address[\inst{2}]{Institute of Physics ASCR, Cukrovarnick\'{a} 10, 162 53
Praha 6, Czech Republic}
\author[Gallagher]{B.L. Gallagher\inst{1}}
\begin{abstract}
Exhibiting antiferromagnetic interlayer coupling in dilute magnetic semiconductor multilayers is essential for the realisation of magnetoresistances analogous to giant magnetoresistance in metallic multilayer structures. In this work we use a mean-field theory of carrier induced ferromagnetism to explore possible (Ga,Mn)As based multilayer structures that might yield antiferromagnetic coupling.
\end{abstract}
\maketitle                   






\section{Introduction}

The exciting new prospect of spin based electronics, known as spintronics, was initiated in 1988 with the discovery of giant magnetoresistance (GMR) in metallic multilayer structures \cite{Baibich:1988_a,Velu:1988_a}.
These structures consist of interposed ferromagnetic (FM) and non-FM layers.
When adjacent FM layers are aligned in antiparallel directions, enhanced spin scattering of carriers causes an increased electrical resistance through the layers.
Dilute magnetic semiconductors (DMS) such as (Ga,Mn)As are a novel class of FM materials which show many spintronic functionalities, and are considered promising candidates for future spintronic applications \cite{Dietl:2000_a}.

Antiferromagnetic (AFM) interlayer exchange coupling (IEC) in DMS based superlattices was theoretically predicted in 1999 using a {\bf \textit {k $\cdot$ p}} kinetic exchange model for carrier mediated ferromagnetism \cite{Jungwirth:1999_a}. This approach considers delocalised charge and adds extra modulation induced by spin-polarised effects. A large magnetoresisticance (MR) was predicited due to the large difference in miniband dispersal between ferromagnetically and antiferromagnetically aligned layers. Recently, IEC has been further explored using a tight binding model \cite{Sankowski:2005_a}. This complementary microscopic approach, although not self-consistant, takes into account atomic orbitals for all the constituent atoms, leading to accurate descriptions of the band structure.
Despite the different approaches used, both methods describe a Ruderman-Kittel-Kasuya-Yosida (RKKY) $2 d k_F$ interlayer coupling mechanism.

In order to realise in DMS materials a phenomenon analagous to GMR in metals, but with potentially a much greater MR ratio, it is essential that AFM interlayer coupling is obtained. However, experimental work into (Ga,Mn)As based multilayer structures has only lead to reports of FM IEC \cite{Kepa:2001_a,Chung:2004_a}. This paper will describe part of a comprehensive review of the multidimensional parameter space available in DMS multilayer systems in order to identify optimal parameters for realising an antiferromagnetically coupled systems.

\section{Theoretical Modelling}

The following calculations will be based on the mean-field theory of carrier induced ferromagnetism \cite{Jungwirth:1999_a}. In this model the band structure is solved using a Kohn-Luttinger kinetic exchange Hamiltonian using a parabolic band {\bf \textit {k $\cdot$ p}} approximation with an additional term $J_{pd}$ representing the $p-d$ exchange interaction between Mn spins and hole spins. 
The limitation of this approach is that a single parabolic band approximation is used, sacrificing full quantitative accuracy for qualitative descriptions of a wide range of systems. Subtleties of the band-structure and spin-orbit effects are neglected. However, qualitative agreement with the data published in \cite{Sankowski:2005_a} at least partially justifies this approach.

In these calculations we use experimentally determined value of $J_{pd} = 55$ meV nm$^3$ \cite{Sinova:2004_c}.
A hole mass $m^* = 0.5 m_e$ and a spin of local Mn moments $S = \frac{5}{2}$ at $T = 0$ K. Thermodynamics is treated on a mean field level. This is done using the standard formalisation of the local-spin density approximation using the Kohn-Sham equations for inhomogeneous systems.

For the calculated IEC energy, $E_c$, positive values correspond to FM interlayer coupling to be energetically favourable, and negative values correspond to AFM interlayer coupling being the favoured configuration.

\section{Results}

 \subsection{GaAs spacer}

\begin{figure*}[tb]
\centering
  \subfigure[]
    {
    \label{Doped_spacer_2m_2pc}
    \includegraphics[width=0.48\textwidth]{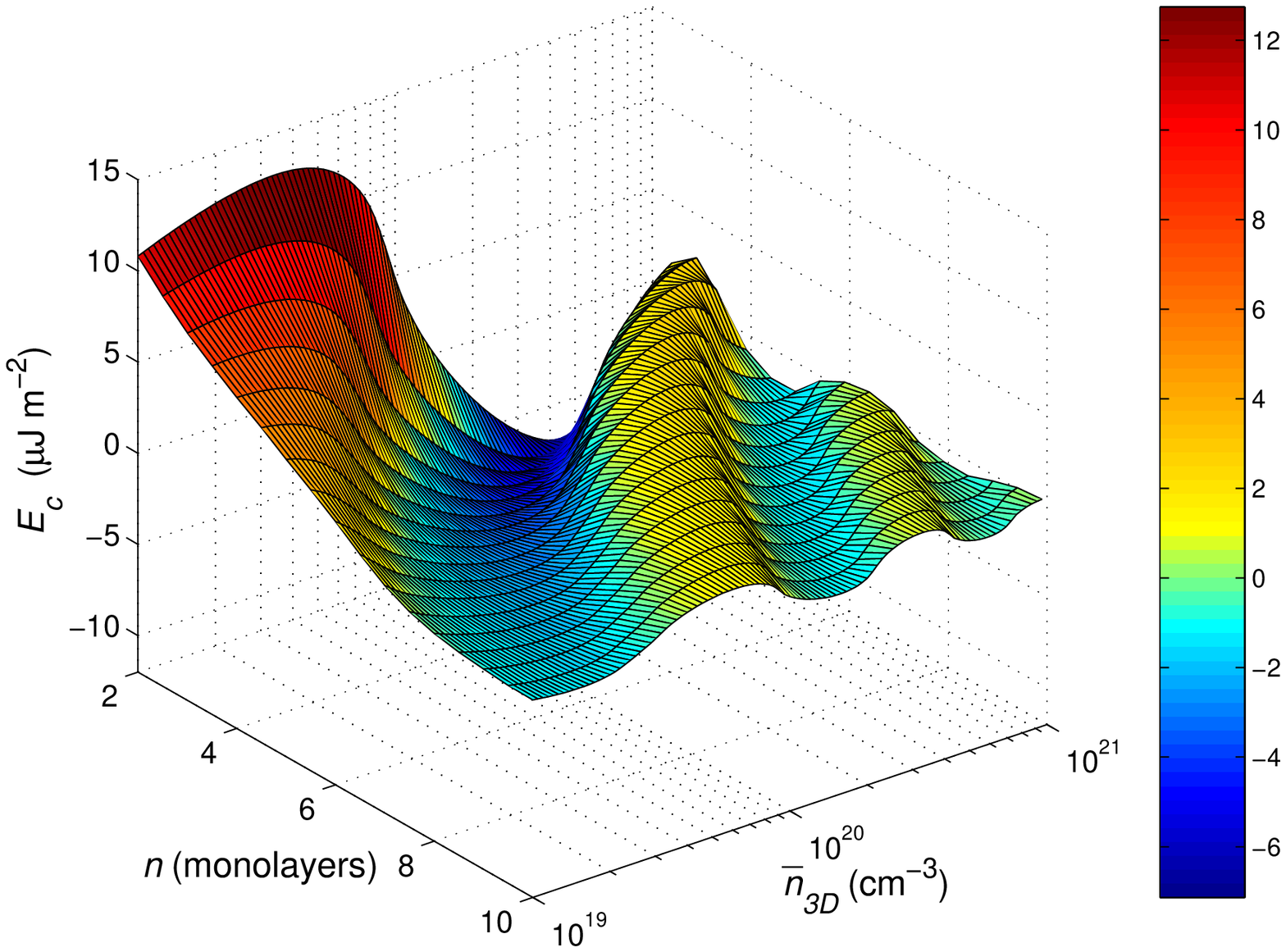}
    }
  \subfigure[]
    {
    \label{Doped_2m_2pc_RKKY}
    \includegraphics[width=0.48\textwidth]{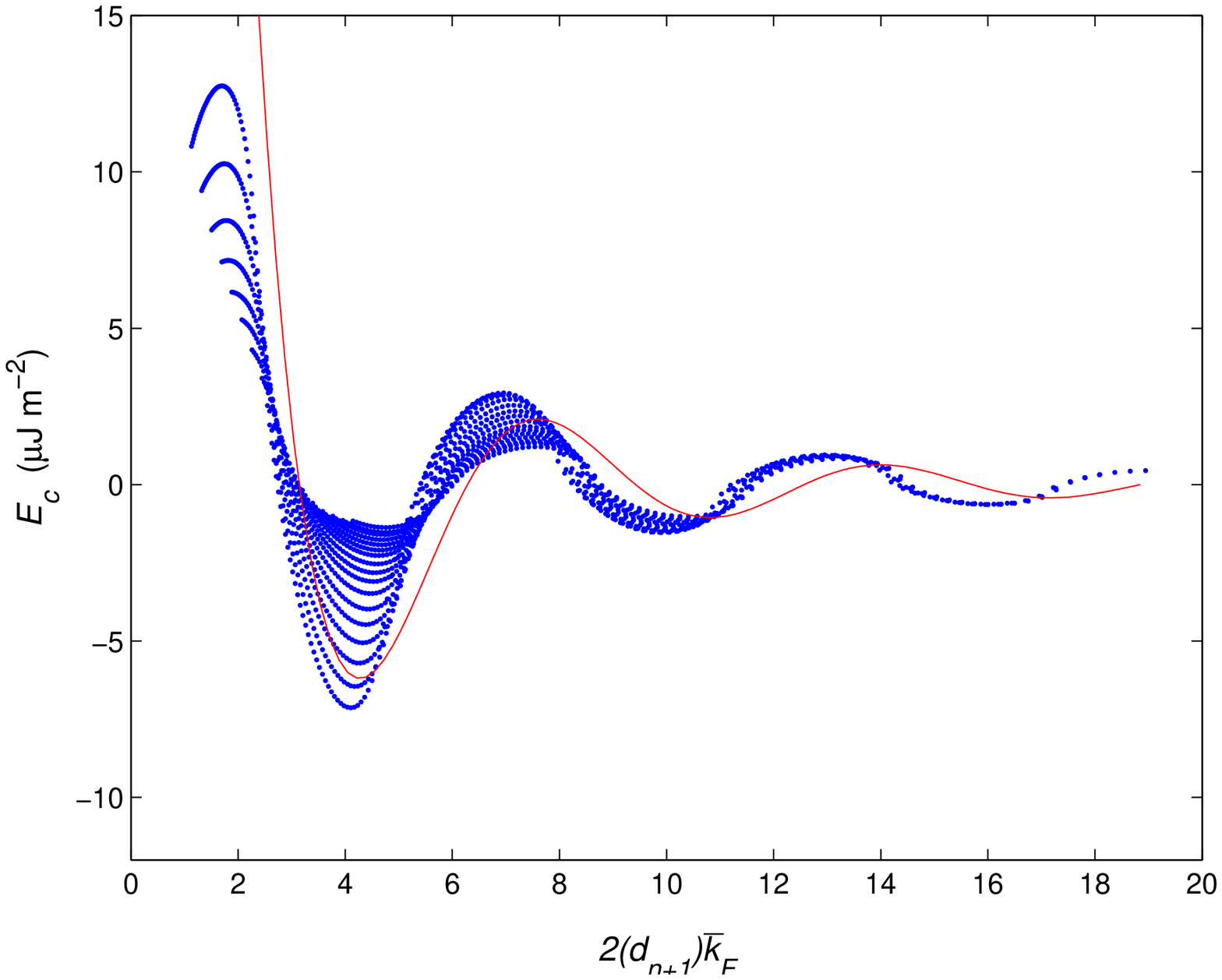}
    }
\caption{
The IEC $E_c$ for a multilayer where the magnetic layers are 2 monolayers thick ($m = 2$) and have a 2\% Mn concentration. There is a uniform impurity concentration throughout the structure. (a) This is shown as a function of the average 3D carrier concentration, $\bar{n}_{3D}$, and the number of monolayers of non-magnetic layer, $n$. (b) Shown as a function of $2 (d_{n + 1}) \bar{k}_F$. The red curve is an estimate of the ideal RKKY range function.}
\label{Doped_spacer}
\end{figure*}

In the RKKY model of interlayer exchange the oscillations occurs as a function of $d k_F$, where $d$ is the separation between the two-dimensionsal magnetic planes and $k_F$ is the Fermi wave vector \cite{Yafet:1987_a}.
In our model we shall denote $d_n$ as the width of the non-magnetic layers, corresponding to $d$ from the RKKY model, and $d_m$ as the width of the magnetic layers. We shall also define the average Fermi wave vector $\bar{k}_F$ as

\begin{equation}
\bar{k}_F = (3 \pi^2 \bar{n}_{3D})^{\frac{1}{3}} ,
\end{equation}

corresponding to the Fermi vector $k_F$ in the ideal RKKY model with a parabolic band. The average three-dimensional (3D) carrier concentration $\bar{n}_{3D}$ is defined as

\begin{equation}
\bar{n}_{3D} = \frac{1}{d_{n+m}} \int_{unit~cell} n_{3D}(z) dz = \frac{n_{2D}}{d_{n+m}} .
\end{equation}
First we shall consider a multilayer structure close to the RKKY limit of infinitely thin magnetic layers surrounded by free unpolarised carriers. So, we shall use thin magnetic layers and a low magnetic moment concentration. In Fig.~\ref{Doped_spacer_2m_2pc} the IEC energy, $E_c$, is plotted against the 3D carrier concentration, $\bar{n}_{3D}$, and the number of monolayers of GaAs in the non-magnetic spacer, $n$, where one monolayer has a thickness of 0.283 nm. The magnetic (Ga,Mn)As layer is 2 monolayers thick and contains 2\% Mn local moment doping. There is a uniform acceptor density throughout the structure which gives an average hole concentration of $4.43 \times 10^{20}$ cm$^{-3}$. In this case there are oscillations as a function of both parameters, analogous to the $d k_F$ oscillations in the ideal quasi one-dimensional RKKY model.

The data from Fig.~\ref{Doped_spacer_2m_2pc} is replotted in Fig.~\ref{Doped_2m_2pc_RKKY} as a function of $2 (d_{n + 1}) \bar{k}_F$. Also plotted is the function

\begin{equation}
y = \alpha \frac{sin(x)}{x^2} ,
\end{equation}

where $\alpha$ is a scaling factor. This function is the asymptotic limit of the pseudo one-dimensional (1D) RKKY range function \cite{Yafet:1987_a}. The strength of the interaction is expected to scale with the density of states, and in the 1D case $\alpha \sim k_F^2$ \cite{Dietl:1997_a}. The different series of points on the graph correspond to the series of different $n$ values from the original plot. For a given $2 (d_{n + 1}) \bar{k}_F$, the points with the largest magnitude are those with the greatest $k_F$; this behaviour is consistent with the expected scaling of  $\alpha$ with $k_F$. The important point to note is the fact that, in order to have improved alignment of the curves, the oscillations were plotted using the parameter $d_{n + 1}$; the non-magnetic spacer thickness plus an additional monolayer. This qualitatively obtained correction factor is necessary because the magnetic layers are not longer 2D planes, so small spacers have proportionally less effect on the centre-to-centre distance of the magnetic layers.

\subsection{AlGaAs spacer}

\begin{figure*}[tb]
\centering
  \subfigure[~$m = 2$, 2\% Mn concentration]
    {
    \label{AlGaAs_spacer_2m_2pc}
    \includegraphics[width=0.48\textwidth]{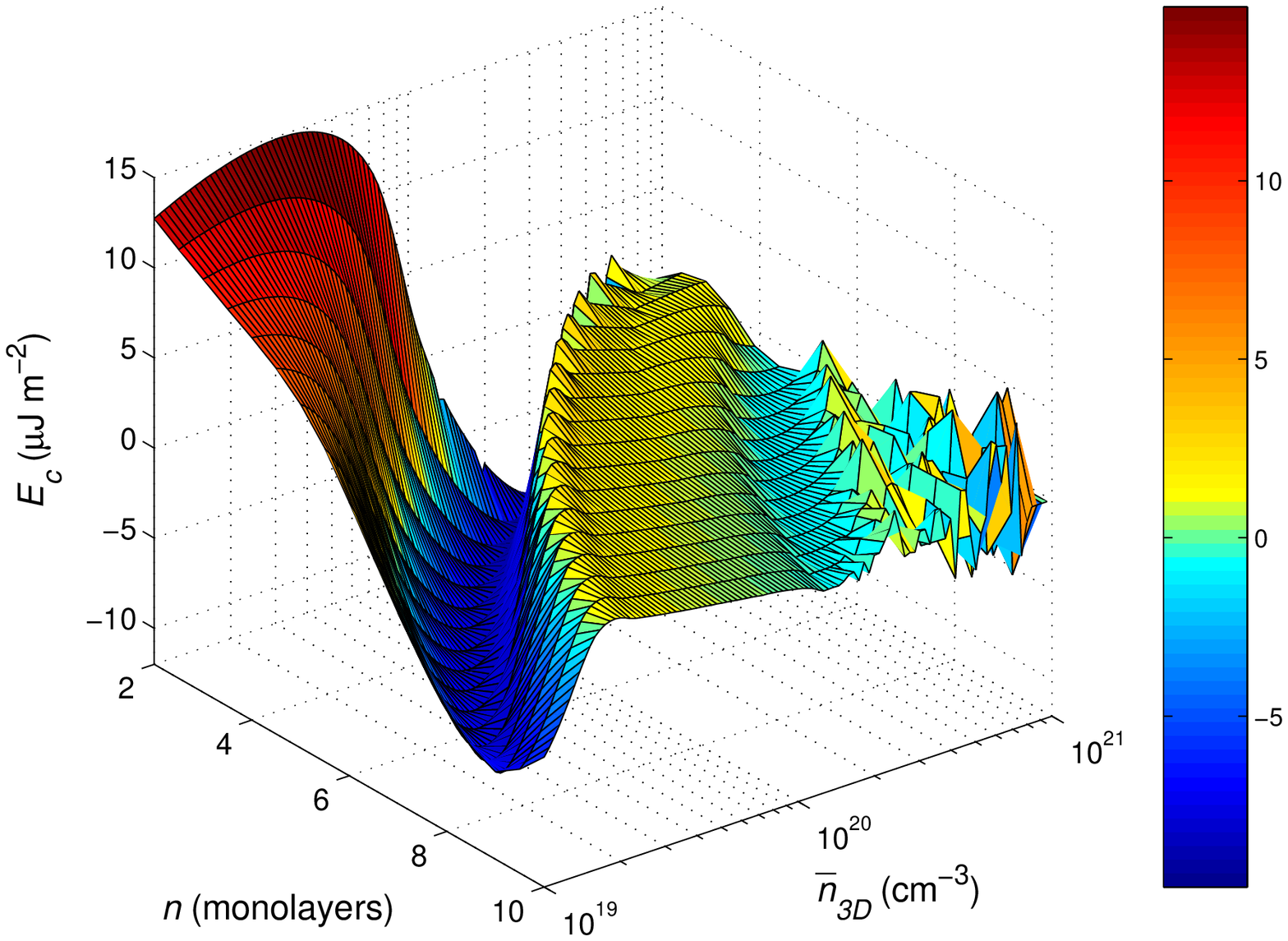}
    }
  \subfigure[~$m = 8$, 2\% Mn concentration]
    {
    \label{AlGaAs_spacer_8m_2pc}
    \includegraphics[width=0.48\textwidth]{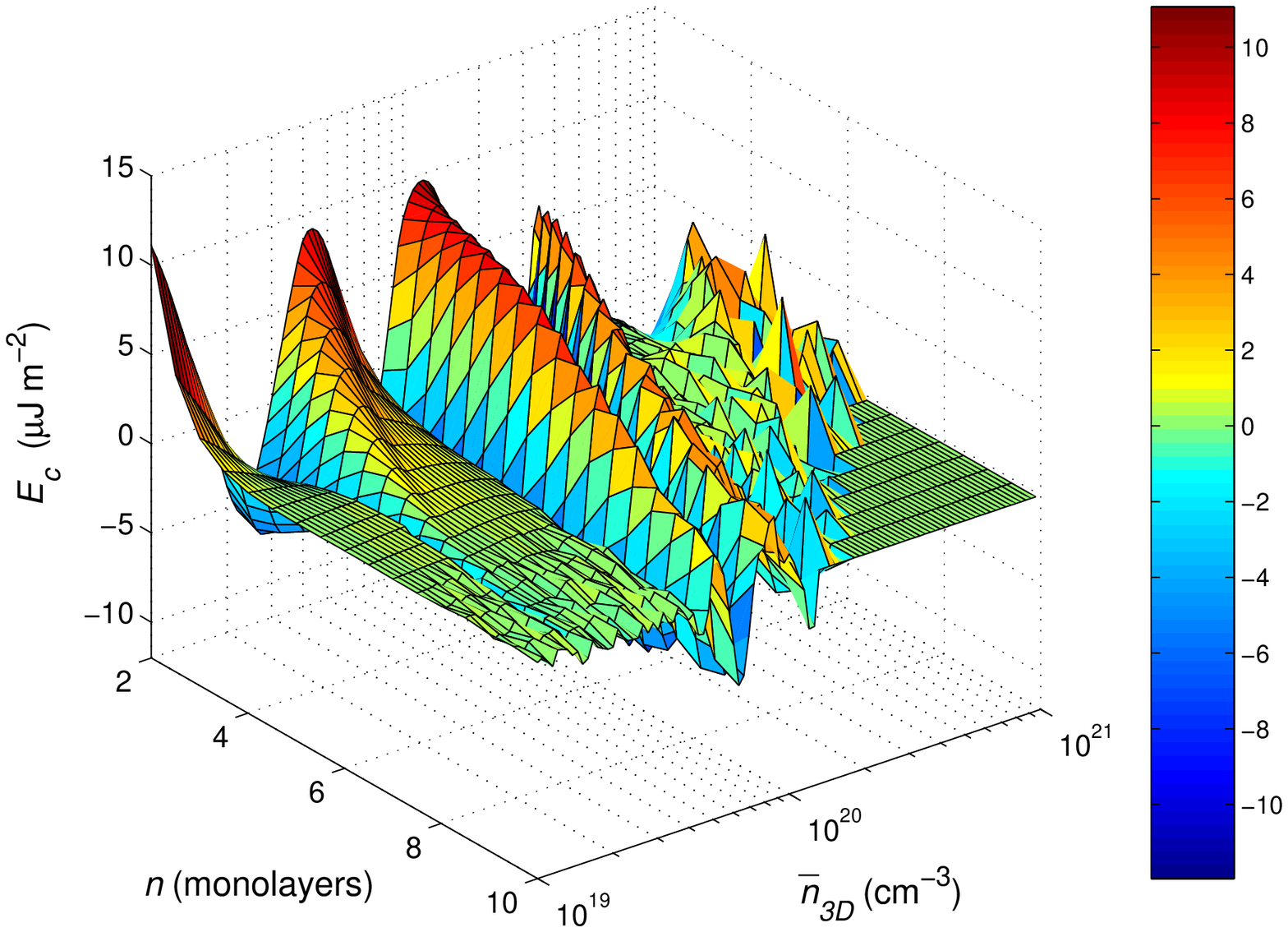}
    }
  \caption{The IEC $E_c$ as a function of the average 3D carrier concentration, $\bar{n}_{3D}$, and the number of monolayers of non-magnetic layer, $n$. The non magnetic layers are Al$_{0.3}$Ga$_{0.7}$As.}
  \label{AlGaAs_spacer}
\end{figure*}

To consider structures further from the the RKKY ideal, a band offset of 150 meV in the valence band is introduced. This represents a non-magnetic layer material of Al$_{0.3}$Ga$_{0.7}$As and will cause a depletion of carriers from the non-magnetic layers, confining them to the magnetic layers.
Fig.~\ref{AlGaAs_spacer_2m_2pc} shows the IEC for a structure with a (Ga,Mn)As magnetic layer of 2 monolayers and an Al$_{0.3}$Ga$_{0.7}$As non-magnetic layer with no acceptor doping. The peak FM and AFM coupling strengths are now stronger than with doped GaAs spacers seen in Fig.~\ref{Doped_spacer_2m_2pc}. However, the $2 d_n \bar{k}_F$ oscillations are damped more rapidly than with the GaAs spacer, resulting in the second FM and AFM peaks being very weak. This additional damping occurs more rapidly with increasing carrier density, $\bar{n}_{3D}$, than with with spacer thickness. As a result, the first AFM peak, which occurs at lower carrier concentrations for greater non-magnetic layer thinkness, is almost constant in value. This is in stark contrast to the GaAs barrier case, where the largest AF coupling when $n = 10$ is less than a quarter of the magnitude as when $n = 2$. This suggests there may be ways to retain strong AFM IEC at greater spacer thicknesses. However, in this case the carrier concentrations at which this occurs would be too low to expect (Ga,Mn)As to be FM.

With high magnetic layer thicknesses the RKKY type oscillations have almost completely disappeared. Fig.~\ref{AlGaAs_spacer_8m_2pc} shows this for $m = 8$, with 2\% Mn doping. In these cases oscillations occur almost exclusively with hole density, being almost independent of the spacer thickness. Furthermore these oscillations have lost their regular character and appear to have a beating type behaviour. This configuration offers some interesting possibilities. Since the oscillations are no longer strongly dependent on spacer thickness, the $\bar{n}_{3D}$ at which the peak AFM IEC occurs for a given $d_n$ is broadly constant. This is particularly interesting as a carrier concentration around $10^{20}$ cm$^{-3}$ is a technologically feasible quantity. It is worth noting that the magnetic layer thicknesses used in the experimental studies are many times thicker, at 25 and 50 monolayers \cite{Kepa:2001_a} and 20 nm ($\sim 70$ monolayers) respectively \cite{Chung:2004_a}; such large (Ga,Mn)As layers would not be expected to yield clear AFM IEC. 

\section{Conclusion}

The composition and structure of (Ga,Mn)As based multilayers can have profound effects on the expected IEC. By examining possible compositions within the broad parameter space that these structures offer it is possible to identify different recipes for devices that might offer the possibility of demonstrating AFM interlayer coupling.

\begin{acknowledgement}
The authors would like to acknowledge the use of the Jupiter supercomputer at the University of Nottingham.
This work was supported by EC through Grant No. FP6-2002-IST-015728, Grant Agency of the Czech Republic under Grant No. 202/05/0575, by Academy of Sciences of the Czech Republic and its Grant Agency under Institutional Support No.AV0Z10100521 and Grant No. A100100530, by the Ministry of Education of the Czech Republic Center for Fundamental Research LC510, and by the UK EPSRC under Grant No. GR/S81407/01.
\end{acknowledgement}

\end{document}